\documentclass[sigconf,screen]{acmart}

\usepackage{booktabs} 
\usepackage[flushleft]{threeparttable}
\usepackage{xcolor}
\usepackage[utf8]{inputenc}
\usepackage{float}

\usepackage{amssymb}
\usepackage{amsmath}
\usepackage{ifthen}
\usepackage{caption}
\usepackage{hyperref}
\usepackage{array,graphicx}
\usepackage{soul}
\usepackage{balance}
\usepackage{pifont}
\usepackage{nicefrac}
\usepackage{mathtools}
\usepackage{tabularx}
\usepackage{xcolor,pifont}
\usepackage{tikz}
\usepackage{svg}
\usepackage{pdfpages}
\usepackage{array}
\usepackage{subcaption}
\usepackage{microtype}
\usepackage{sparklines}
\usepackage{booktabs}
\usepackage{graphicx}
\usepackage{longtable}
\usepackage{array}
\usepackage{xcolor}
\usepackage{supertabular}
\usepackage{booktabs}
\usepackage{xcolor}
\usepackage{adjustbox}
\usepackage{enumitem}

\newcommand*\colourcheck[1]{%
  \expandafter\newcommand\csname #1check\endcsname{\textcolor{#1}{\ding{52}}}%
}
\newcommand*\colourcross[1]{%
  \expandafter\newcommand\csname #1cross\endcsname{\textcolor{#1}{\ding{56}}}%
}
\usepackage[vlined, boxruled, linesnumbered] {algorithm2e}
\SetKw{KwBy}{by}
\SetKw{KwBreak}{break}
\SetKw{KwReturn}{return}
\def\HiLi{\leavevmode\rlap{\hbox to \hsize{\color{gray!35}\leaders\hrule height .8\baselineskip depth .5ex\hfill}}}

\newcommand{\tool}{Cam2Sim\xspace}
\newcommand{\head}[1]{\noindent\textbf{#1.}}

\newlength\BARWIDTH
\newlength\BARHEIGHT
\newlength\BARWIDTHfour
\SetKwComment{Comment}{$\triangleright$\ }{}
\SetCommentSty{itshape}
\newboolean{showcomments}
\setboolean{showcomments}{true}
\ifthenelse{\boolean{showcomments}}
{\newcommand{\nb}[2] {
  \fcolorbox{black}{gray!20}{\bfseries\sffamily\scriptsize#1:}
  {\sf\small$\blacktriangleright$\textit{#2}$\blacktriangleleft$}
}
}
{\newcommand{\nb}[2]{}
}

\newcounter{fcounter}
\newcommand{\curl}[1]{\footnote{\url{#1}}}

\usepackage{tikz}

\expandafter\def\csname rqthreecolor@success\endcsname{green!70!black}
\expandafter\def\csname rqthreecolor@skip\endcsname{blue!70!black}
\expandafter\def\csname rqthreecolor@fail\endcsname{red!80!black}

\newcommand{\RQThreeColor}[1]{\csname rqthreecolor@#1\endcsname}

\makeatletter
\newcommand{\thickhline}{%
    \noalign {\ifnum 0=`}\fi \hrule height 1pt
    \futurelet \reserved@a \@xhline
}
\makeatother

\setcopyright{cc}
\setcctype{by}
\acmDOI{10.1145/3832783.3834639}
\acmYear{2026}
\copyrightyear{2026}
\acmISBN{979-8-4007-2882-2/2026/10}
\acmConference[ASE '26]{Proceedings of the 41st IEEE/ACM International Conference on Automated Software Engineering}{October 12--16, 2026}{Munich, Germany}
\acmBooktitle{Proceedings of the 41st IEEE/ACM International Conference on Automated Software Engineering (ASE '26), October 12--16, 2026, Munich, Germany}
\acmSubmissionID{ase26tool-p120-p}
\received{2026-05-12}
\received[accepted]{2026-06-19}

\begin{document}
\title{Cam2Sim: Neural Scenario Reconstruction for Closed-Loop Autonomous Driving Simulation}

\author{Davide Jannussi}
\orcid{0009-0000-2200-9872}
\affiliation{%
  \institution{Politecnico di Torino}
  \city{Torino}
  \country{Italy}
}
\email{s331391@studenti.polito.it}

\author{Stefano Carlo Lambertenghi}
\orcid{0009-0007-0513-1877}
\affiliation{%
  \institution{TU Munich, fortiss}
  \city{Munich}
  \country{Germany}
}
\email{stefanocarlo.lambertenghi@tum.de}

\author{Constantin Carste}
\orcid{0009-0002-0612-7064}
\affiliation{%
  \institution{TU Munich}
  \city{Munich}
  \country{Germany}
}
\email{constantin.carste@tum.de}

\author{Andrea Stocco}
\orcid{0000-0001-8956-3894}
\affiliation{%
  \institution{TU Munich, fortiss}
  \city{Munich}
  \country{Germany}
}
\email{andrea.stocco@tum.de}

\begin{abstract}
Simulation-based testing enables safe and repeatable evaluation of autonomous driving systems, but its effectiveness is limited by the gap between synthetic simulator outputs and real-world camera observations. We present \tool, a tool that transforms real-world driving recordings into executable CARLA scenarios. Starting from camera images and poses, \tool reconstructs road geometry, ego trajectories, parked vehicles, and simulation assets, and augments the environment with Gaussian Splatting to render camera observations resembling the original recording. The framework supports ROS-based data extraction, parked-vehicle detection, OpenStreetMap-based map generation, CARLA scenario construction, Gaussian Splatting training, trajectory replay, and closed-loop testing.
We validate \tool on a real urban-driving scenario with an end-to-end driving model, evaluating reconstruction quality, image-generation quality, and closed-loop behavior against a simulation-only baseline and the real-world reference. Results show that Gaussian-Splatting-based rendering reduces the visual gap with respect to simulator rendering and improves behavioral similarity to the real-world reference.
The artifact is publicly available at \url{https://github.com/ast-fortiss-tum/cam2sim}, and a screencast is available at \url{https://youtu.be/KmZ74l1__lI}.
\end{abstract}


\maketitle

\section{Introduction}\label{sec:introduction}

Autonomous driving systems (ADS)~must be evaluated before deployment because failures can have safety-critical consequences~\cite{zhong2021survey}. 
Modern ADS rely heavily on machine learning and deep learning components for perception and decision-making~\cite{2020-Riccio-EMSE}, whose data-driven nature introduces failure modes that differ fundamentally from those of traditional software, motivating specialized testing techniques~\cite{humbatova2020taxonomy}.
Real-world testing provides the strongest evidence of system performance under realistic operating conditions, but it is expensive, difficult to reproduce, time-consuming, and unsafe for many corner cases~\cite{2023-Stocco-TSE}.

Dataset-based evaluation improves reproducibility, yet it replays fixed observations and therefore cannot capture the closed-loop interaction between the ADS and its environment~\cite{2023-Stocco-EMSE}. As a result, failures emerging from the interaction between perception, planning, and control may remain undetected. 

Simulation-based testing addresses this limitation by enabling the ADS to interact with a controllable and reproducible virtual environment. However, its effectiveness is limited by the sim-to-real gap~\cite{2023-Stocco-TSE}. Differences in textures, illumination, scene appearance, and other modeling inaccuracies can alter the input distribution perceived by the ADS, leading to behaviors that deviate from those observed in the real world~\cite{2025-lambertenghi-ASE,2024-Lambertenghi-ICST}. Consequently, improving the realism and trustworthiness of simulation environments has become a central objective of recent research in ADS testing~\cite{sorokin2025simulatorensemblestrustworthyautonomous,biagiola2023better}.

Recent advances in neural rendering and scene reconstruction, including GANs, diffusion models, Neural Radiance Fields, and Gaussian Splatting, have shown promise in generating visually realistic scenes from real-world recordings~\cite{2024-Lambertenghi-ICST,Zhu2024SceneRT,2026-Guo-OJ-ITS}. However, existing approaches such as DriveRecon~\cite{lu2025drivingrecon} mainly focus on rendering viewpoints or image sequences rather than executable, closed-loop simulations~\cite{2025-Baresi-ICSE}. Conversely, prior work on scenario reconstruction from police reports and structured accident descriptions focuses on rebuilding executable scenarios~\cite{gambi-reconstruction}, but does not address the perception-level sim-to-real gap. 

To address this gap, we present \tool, the first framework, to the best of our knowledge, that transforms real-world 2D driving recordings and camera datasets into 3D executable CARLA test scenarios for closed-loop ADS testing. Starting from front-facing camera images and poses, \tool reconstructs road geometry, ego trajectories, parked vehicles, and simulation assets inside CARLA~\cite{carla}. The framework then augments the reconstructed environment with Gaussian Splatting~\cite{3gs} to render camera observations that resemble the original recording. The resulting scenarios support both trajectory replay and closed-loop ADS testing, where the ADS receives the GS-rendered camera observations during execution.
\tool supports ROS-based extraction, parked-vehicle detection from camera or LiDAR data, OpenStreetMap-based map generation, CARLA scenario construction, local GS training, trajectory replay, and closed-loop testing with a camera-based ADS. It outputs CARLA-ready maps, trajectories, parked-vehicle placements, GS models, replay videos, steering logs, and execution traces for downstream analysis.

We evaluate \tool on a real-world urban-driving scenario using a camera-based end-to-end driving model. Our evaluation assesses reconstruction quality and closed-loop behavioral fidelity under simulator-rendered and GS-rendered observations. Results show that GS-rendered observations reduce the visual discrepancy with respect to standard simulator rendering while enabling behaviors that more closely resemble the real-world reference runs.

This tool paper makes the following contributions:

\head{Tool}
We present \tool, a framework that transforms real-world recordings into executable CARLA scenarios for closed-loop ADS testing through reconstruction and Gaussian-Splatting-based rendering in CARLA.

\head{Evaluation in a real-world urban driving scenario}
We demonstrate improvements in reconstruction quality and behavioral similarity compared to a simulation-only baseline. In our experiments, the GS-augmented simulation successfully completes all runs, while the simulation-only baseline consistently fails the scenario.

\begin{figure*}[t]
    \centering
    \includegraphics[width=1\linewidth]{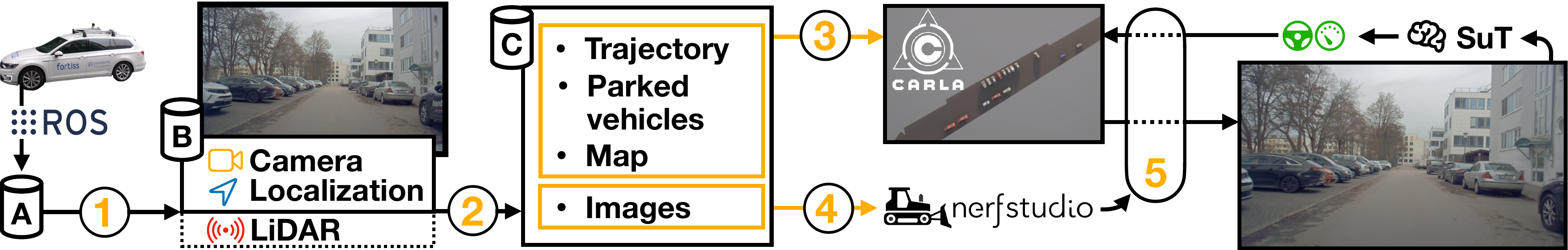}
    \caption{Overview of \tool: (1)~Data Extraction, (2)~Dataset Processing, (3)~Simulation Data Generation, (4)~Gaussian Splatting Preparation, and (5)~Driving Simulation.}
    \label{fig:schema}
\end{figure*}
\section{\tool Framework}\label{sec:schema}

\autoref{fig:schema} shows the \tool workflow, which consists of five components: (1)~data extraction, (2)~dataset processing, (3)~simulation-asset generation, (4)~Gaussian Splatting preparation, and (5)~driving simulation. Together, these components transform raw recordings or prepared camera datasets into CARLA-ready scenarios with map data, trajectories, parked-vehicle positions, GS-training inputs, and executable replay or closed-loop simulations.

\subsection{Data Extraction}\label{subsec:data-extraction}

The data extraction component converts ROS-based recordings into the structured dataset format used by \tool. It can be skipped when front-facing camera images and corresponding camera poses are already available. The minimum required input is a sequence of camera images and poses, which define the camera trajectory and are sufficient for scenario reconstruction, GS training, and replay.

When ROS bags are used, \tool extracts camera frames and associates each frame with the ego pose at the corresponding timestamp. It can also extract optional LiDAR, odometry, steering angles, and model-prediction streams. These signals improve reconstruction and support validation: LiDAR enables more accurate parked-vehicle detection, while steering and model-prediction logs support comparison with the ADS. Sensor streams with different frequencies are synchronized with ego poses through timestamp-based interpolation. The output is a structured dataset containing camera frames, poses, and point clouds, and trajectories.

\subsection{Dataset Processing}\label{subsec:dataset-processing}

The dataset processing component extracts the information required for scenario reconstruction, map generation, GS training, and simulation. The camera trajectory is used as the reference path for reconstructing and replaying the scenario.

\tool detects parked vehicles and provides a graphical interface for manual refinement. From camera data, it applies monocular 3D object detection with FCOS3D~\cite{fcos3d}, transforms detections into world coordinates using camera-to-ego calibration and timestamp-aligned ego poses, and clusters repeated detections across frames. Final positions are computed as confidence-weighted averages~\cite{2021-Stocco-JSEP}, while lane-side and orientation labels are obtained through majority voting. When LiDAR is available, \tool can instead detect parked vehicles with PointPillars~\cite{pointpillars}; users can then move, rotate, insert, or delete boxes to produce refined parked-vehicle files.

This component also prepares map and GS inputs. From the trajectory or a manually specified address, \tool retrieves OpenStreetMap data through the Overpass API~\cite{overpass}. For GS training, it selects frames at a configurable interval, removes the visible ego-vehicle hood from the selected frames, generates sky masks using SegFormer trained on Cityscapes~\cite{segformer,cityscapes}, and splits the route into overlapping chunks containing images, masks, and pose metadata.

\subsection{Simulation Data Generation}\label{subsec:simulation-data-generation}

The simulation-data generation component converts the processed scenario into CARLA-ready assets: the OpenDRIVE map, replay trajectory, initial hero-vehicle pose, and parked-vehicle spawn positions. \tool exports both center-position and rear-axle trajectory variants to account for the difference between recorded sensor poses and CARLA vehicle transforms.

Coordinate conversion maps dataset positions to CARLA coordinates through the OpenDRIVE map. When geographic coordinates are used, \tool converts WGS84 coordinates into the OpenDRIVE reference frame and adapts the axis convention for CARLA. The tool also provides utilities to load the generated map, inspect trajectory alignment and parked-vehicle placement, and prepare the final CARLA scenario.

\subsection{Gaussian Splatting Preparation}\label{subsec:gs-preparation}
Using the processed frames and reconstructed trajectories, the GS preparation component trains models used to render camera observations resembling the original recording.

\tool runs COLMAP~\cite{colmap} with the selected images, calibrated camera model, and optional sky masks to estimate camera poses and a sparse 3D point cloud. It then trains GS models with Nerfstudio~\cite{nerfstudio}, using \texttt{splatfacto} or \texttt{splatfacto-big}; sky masks exclude sky regions so the model focuses on static scene content. After training, \tool aligns the Nerfstudio and dataset coordinate systems by matching reconstructed camera poses with the original pose annotations and estimating the corresponding alignment parameters. These models are later used to render GS images from CARLA camera poses.

Because the deployed ADS directly consumes the rendered camera stream, successful closed-loop execution also provides an operational assessment of the generated observations: severe rendering artifacts or missing scene information would directly affect steering behavior and trajectory stability.

\begin{figure*}[t]
    \centering
    \includegraphics[width=\linewidth]{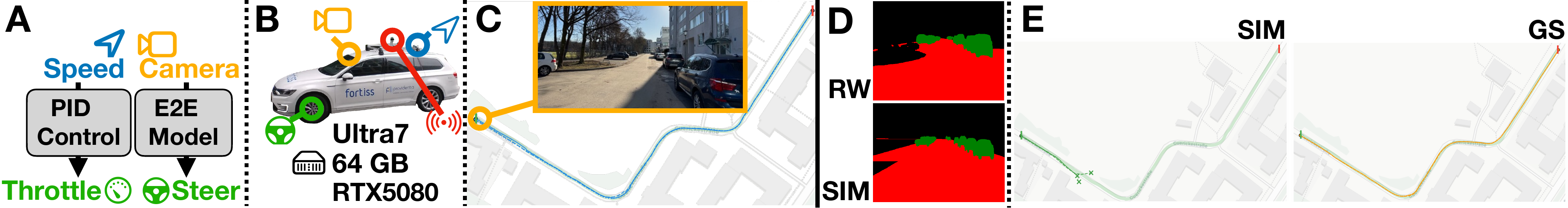}
    \caption{Validation setup and results: (A)~the ADS, (B)~the autonomous-driving platform used to collect the target recording, (C)~the real-world collection scenario, (D)~results of Scenario Reconstruction, and (E)~Behavior Fidelity.}
    \label{fig:validation}
\end{figure*}

\subsection{Driving Simulation}\label{subsec:driving-simulation}

The driving simulation component executes the reconstructed scenario in CARLA using the generated map, trajectories, parked vehicles, camera configuration, and, when enabled, the trained GS models and alignment files. \tool supports both trajectory replay and closed-loop execution.

In trajectory-replay mode, the hero vehicle follows the recorded trajectory frame by frame. This mode is used to inspect the reconstructed scenario and compare observations from the same viewpoints as the original recording. \tool can save CARLA-rendered images, GS-rendered images, and side-by-side outputs.

In closed-loop mode, the ADS controls the ego vehicle. \tool currently supports a camera-based DAVE-2 model~\cite{bojarski2016end}, served by a separate inference process that predicts steering commands from RGB images. The ADS can receive either raw CARLA images, forming the simulation-only baseline, or GS-rendered images to evaluate whether GS-based rendering reduces the gap to the original driving scenario. The output includes replayed camera streams, trajectories, steering commands, and execution logs for evaluating reconstruction quality, visual fidelity, and behavior similarity.



\section{Preliminary Validation}\label{sec:empirical-study}

We evaluate whether \tool can transform a real-world driving recording into a reconstructed simulation that is geometrically consistent with the original scenario, visually closer to the original camera stream, and suitable for closed-loop ADS testing. The validation follows the structure of the pipeline. First, we assess whether the reconstruction and simulation-generation components correctly reproduce the road geometry, ego trajectory, and parked vehicles of the target scenario. Second, we evaluate whether Gaussian Splatting preserves visually relevant scene information required for closed-loop ADS execution.

To this aim, we conduct a real-world data-collection campaign using an ADS-equipped research vehicle on a public urban road and use the resulting recording as the target scenario for \tool. The evaluation compares three domains: (i)~the real-world reference execution, (ii)~a simulation-only baseline using the reconstructed CARLA scenario, and (iii)~a GS-augmented simulation in which camera observations are rendered using Gaussian Splatting.

\head{System Under Test}
We use a camera-based end-to-end DAVE-2 model~\cite{bojarski2016end} that predicts continuous steering directly from RGB images captured by the front-facing camera. During execution, longitudinal control is handled independently by a PID controller that maintains a fixed speed of 20~km/h (\autoref{fig:validation}~(A)).

\head{Autonomous Driving Platform}
We use \textit{fortuna}~\cite{buechel2019fortuna_preprint}, a modified Volkswagen Passat Variant GTE research vehicle equipped with cameras, Velodyne LiDARs, and GNSS/INS localization. The front-facing camera is used both for scenario reconstruction and as input to the ADS, while the roof-mounted LiDAR supports parked-vehicle detection during real-world data collection (\autoref{fig:validation}~(B)).

\head{Collection Scenario}
Experiments are conducted in Munich on a 450~m two-way residential street with a 30~km/h speed limit. The scenario includes an unmarked roadway, sidewalks, three bends, and parked vehicles partially narrowing the driving lane, providing a realistic yet controlled urban evaluation environment (\autoref{fig:validation}~(C)).

\subsection{Scenario Reconstruction Quality}

We first evaluate whether \tool reconstructs the target scenario with sufficient geometric fidelity for simulation-based testing. We apply the reconstruction pipeline to the collected recording and generate a CARLA scenario containing the reconstructed road map, replay trajectory, and parked-vehicle placements. We then execute trajectory replay and compare the rendered semantic maps against the real-world semantic observations from the same viewpoints.

We assess reconstruction quality using Intersection over Union (IoU) over the road, car, and background classes. Across 3,144 replayed frames, \tool achieves a mean IoU (mIoU) of $0.774$ ($\sigma=0.071$). The reconstructed road layout achieves the highest agreement with the real-world reference, with a mean IoU of $0.847$ ($\sigma=0.056$), while background regions achieve $0.889$ ($\sigma=0.037$). Vehicle reconstruction is more challenging due to localization inaccuracies and partial occlusions, yet the car class still achieves a mean IoU of $0.577$ ($\sigma=0.206$). An example semantic comparison is shown in \autoref{fig:validation}~(D). Overall, these results suggest that \tool reconstructs urban-driving scenarios with sufficient geometric fidelity to support replay and closed-loop ADS testing.

\subsection{Image Generation Quality}

We next evaluate whether GS reduces the visual gap between the reconstructed CARLA scenario and the real recording. Starting from the reconstructed scenario, we train the GS model using Step~4 on the prepared camera frames, poses, sky masks, and local training splits. We then execute Step~5 in trajectory-replay mode and render the camera stream with GS from the replayed viewpoints.

Following the evaluation methodology proposed in~\cite{2024-Lambertenghi-ICST}, we assess image quality using both \emph{single-image} and \emph{distribution-level} metrics. Single-image metrics compare corresponding real and generated frames from identical viewpoints, measuring reconstruction fidelity using Peak Signal-to-Noise Ratio (PSNR)~\cite{inception-metric} and perceptual similarity using the Learned Perceptual Image Patch Similarity (LPIPS) metric~\cite{lipsis}. Distribution-level metrics instead compare the distributions of generated and real images, measuring dataset-level visual realism using the Fréchet Inception Distance (FID)~\cite{fid-metric} and the Kernel Inception Distance (KID)~\cite{kid-citations}.

Across all 3,144 replayed viewpoints, GS improves PSNR by 48.8\% (11.23\,dB$\rightarrow$16.72\,dB) and reduces LPIPS by 35.5\% (0.757$\rightarrow$0.488). At the distribution level, GS reduces FID by 84.7\% (245.9$\rightarrow$37.7) and KID by 95.3\% (0.246$\rightarrow$0.012). 
These improvements at both the image and dataset levels demonstrate that GS significantly reduces the visual gap with respect to standard CARLA rendering, producing observations that are closer to the original camera stream.

\subsection{Behavior Fidelity}

Finally, we evaluate whether the GS-augmented simulation improves system-level testing realism. We execute the driving-simulation component in closed-loop mode with the deployed ADS in three domains: the real-world setting, a simulation-only baseline, and a GS-augmented simulation. In the simulation-only baseline, the ADS receives raw CARLA camera images; in the GS-augmented setting, it receives camera observations rendered by the GS model.

To account for variability and the non-determinism of the end-to-end driving model, we execute each domain three times. We compare the generated trajectories against the real-world reference executions using failure and completion rates, as well as trajectory-similarity and steering-smoothness metrics, including Fréchet distance, corridor violations (percentage of trajectory samples outside the driving corridor), lateral excess, and steering jitter.

Results are shown in \autoref{fig:validation}~(E) and Table~\ref{tab:behavior-fidelity}. The simulation-only baseline consistently fails to complete the scenario, whereas the GS-augmented simulation successfully completes all runs. Compared to the simulation-only baseline, the GS-based simulation produces trajectories that are closer to the real-world reference and stable steering behavior across all runs. Overall, these preliminary results suggest that GS-rendered observations improve the behavioral fidelity of closed-loop ADS testing compared to standard simulator rendering.

\begin{table}[t]
\centering
\caption{Behavior fidelity comparison between the real-world execution, the simulation-only baseline, and the GS-augmented simulation. Higher completion rates and lower trajectory-deviation metrics indicate behavior closer to the real-world reference. OR = out of road, CC = car crash.}
\label{tab:behavior-fidelity}
\small
\renewcommand{\arraystretch}{0.97}
\begin{tabular}{lccc}
\toprule
\textbf{Metric} & \textbf{Real} & \textbf{\tool} & \textbf{Sim} \\
\midrule

Failure Rate &
0/3 &
0/3 &
3/3 \\

Completion Rate (\%) &
100--100--100 &
100--100--100 &
21--23--20 \\

Failure Type (OR--CC) &
-- &
-- &
2--1 \\

\midrule
\multicolumn{4}{c}{\textit{Trajectory Similarity}} \\
\midrule

Fréchet Distance (m) &
0.00 &
2.22 &
-- \\

Corridor Violations (\%) &
0.0 &
67.3 &
-- \\

Mean Excess (m) &
0.00 &
0.330 &
-- \\

Excess When Out (m) &
0.00 &
0.491 &
-- \\

\midrule
\multicolumn{4}{c}{\textit{Steering Smoothness}} \\
\midrule

Average Jitter (rad/s) &
1.167 &
0.867 &
-- \\

Maximum Jitter (rad/s) &
6.452 &
4.983 &
-- \\

\bottomrule
\end{tabular}

\end{table}
\section{Conclusions}\label{sec:conclusions}

We presented \tool, a tool that transforms real-world recordings or camera datasets into executable CARLA scenarios augmented with Gaussian Splatting. Starting from camera images and poses, the tool reconstructs scenario geometry, prepares simulation-ready assets, trains local GS models, and supports both trajectory replay and ADS closed-loop execution.
As future work, we plan to extend \tool with roadside elements and dynamic-objects reconstruction, support for additional neural-rendering backends and sensor modalities, and larger-scale evaluations across diverse driving environments and ADS architectures.

\section{Data Availability Statement}

\tool is available as an actively maintained GitHub repository~\cite{replication-package} and a permanently archived Zenodo release~\cite{zenodo}. The artifact includes the complete framework, the validation ROS bag, the camera-based ADS, and scripts to reproduce the Gaussian Splatting and closed-loop evaluations. Validation was performed on the research vehicle using an Intel Core Ultra~9 computer (64~GB RAM, RTX~5080), while Gaussian Splatting training and offline simulation were performed on an Intel Core Ultra~9 workstation (32~GB RAM, RTX~4090). The validation dataset is publicly available in anonymized form, with faces and license plates blurred, enabling reproduction of all reported results with minimal changes.


\balance
\bibliographystyle{ACM-Reference-Format}
\bibliography{paper}

\end{document}